%
\input phyzzx
\catcode`@=11
%
%
\newtoks\UT
\newtoks\monthyear
\Pubnum={UT-\the\UT}
\UT={741}
\monthyear={February, 1996}
\def\p@bblock{\begingroup \tabskip=\hsize minus \hsize
    \baselineskip=1.5\ht\strutbox \topspace-2\baselineskip
    \halign to\hsize{\strut ##\hfil\tabskip=0pt\crcr
    \the\Pubnum\cr\the\monthyear\cr
    }\endgroup}
\def\bftitlestyle#1{\par\begingroup \titleparagraphs
    \iftwelv@\fourteenpoint\else\twelvepoint\fi
    \noindent {\bf #1}\par\endgroup}
\def\title#1{\vskip\frontpageskip \bftitlestyle{#1} \vskip\headskip}
%
%

%
%

%
\def\journal#1&#2(#3){\begingroup \let\journal=\dummyj@urnal
    \unskip, \sl #1\unskip~\bf\ignorespaces #2\rm
    (\afterassignment\j@ur \count255=#3) \endgroup\ignorespaces}
\def\andjournal#1&#2(#3){\begingroup \let\journal=\dummyj@urnal
    \sl #1\unskip~\bf\ignorespaces #2\rm
    (\afterassignment\j@ur \count255=#3) \endgroup\ignorespaces}
\def\andvol&#1(#2){\begingroup \let\journal=\dummyj@urnal
    \bf\ignorespaces #1\rm
    (\afterassignment\j@ur \count255=#2) \endgroup\ignorespaces}

\def\NP{Nucl.~Phys.}
\def\PL{Phys.~Lett.}
\def\PR{Phys.~Rev.}

\def\PTP{Prog.~Theor.~Phys.}

\catcode`@=12
%


\titlepage

\title{Dynamical Supersymmetry Breaking
       in Vector-like Gauge Theories}

\author{Izawa {\twelverm Ken-Iti}
\foot{\rm JSPS Research Fellow.}
{\twelverm and Tsutomu} Yanagida}
\address{Department of Physics, University of Tokyo \break
                    Tokyo 113, Japan}

\abstract{
We provide vector-like gauge theories
which break supersymmetry dynamically.
}

\endpage

\doublespace


\def\l{\lambda}

\def\L{\Lambda}

\def\j{\journal}


\REF\Aff{I.~Affleck, M.~Dine, and N.~Seiberg \j \NP &B256 (85) 557.}

\REF\Sei{N.~Seiberg \j \PR &D49 (94) 6857.}

\REF\Int{N.~Seiberg \j \PL &B318 (93) 469; \nextline
         K.~Intriligator and N.~Seiberg, hep-th/9509066; \nextline
         Izawa K.-I. \j \PTP &95 (96) 199.}

\REF\Ora{P.~Fayet \j \PL &B58 (75) 67; \nextline
         L.~O'Raifeartaigh \j \NP &B96 (75) 331.}

\REF\Wit{E.~Witten \j \NP &B202 (82) 253.}

\REF\Pou{K.~Intriligator and P.~Pouliot \j \PL &B353 (95) 471.}

\REF\Iza{Izawa K.-I. and T.~Yanagida \j \PTP &94 (95) 1105.}

\sequentialequations


There is a piece of folklore which holds that vector-like gauge theories
cannot break supersymmetry dynamically.
In this letter, we point out remarkable exceptions to this piece of folklore.

Let us consider a supersymmetric SU(2) gauge theory with
four doublet chiral superfields $Q_i$.
We also introduce six singlet chiral superfields $Z^{ij} = -Z^{ji}$.
Here $i$ and $j$ denote the flavor indices ($i, j = 1, \cdots, 4$).

The tree-level superpotential of our model is given by
\foot{This tree-level superpotential is natural
since it possesses two global symmetries.
One is an axial U(1) symmetry associated with
a $Q_i$ phase transformation
and the other is an anomaly-free $R$ symmetry.}
$$
  W_{tree} = \l_{ij}^{kl} Z^{ij} Q_k Q_l,
 \eqn\INT
$$
where $\l_{ij}^{kl}$ denote generic coupling constants
with $\l_{ij}^{kl} = -\l_{ji}^{kl} = -\l_{ij}^{lk}$.
The pecuriarity of this superpotential resides in that
it raises all the $D$-flat directions in the doublets $Q_i$,
which is a necessary condition for supersymmetry to break down\rlap.
\refmark{\Aff}
Of course, supersymmetry remains unbroken perturbatively in this model.

The exact effective superpotential of the model, which takes into account
the full nonperturbative effects,
may be written in terms of gauge-invariant low-energy degrees of freedom
\refmark{\Sei}
$$
  V_{ij} = -V_{ji} \sim Q_i Q_j
 \eqn\COMP
$$
as follows:
$$
  W_{eff} = X({\rm Pf} V_{ij} - \L^4) + \l_{ij}^{kl} Z^{ij} V_{kl},
 \eqn\EFF
$$
where $X$ is an additional chiral superfield,
${\rm Pf} V_{ij}$ denotes the Pfaffian of the antisymmetric matrix
$V_{ij}$, and $\L$ is a dynamical scale of the SU(2) gauge interaction\rlap.
\refmark{\Sei, \Int}
This is none other than a superpotential of the O'Raifeartaigh type\rlap.
\refmark{\Ora}
Namely, this effective superpotential yields conditions for
supersymmetric vacua
$$
  {\rm Pf} V_{ij} = \L^4, \quad \l_{ij}^{kl} V_{kl} = 0,
 \eqn\SVAC
$$
which cannot be satisfied simultaneously as far as $\L \neq 0$.
Therefore we conclude that supersymmetry is dynamically broken
in our model.

We note that this conclusion is not in contradiction
with the index argument\rlap.
\refmark{\Wit}
The doublets $Q_i$ cannot be decoupled by means of mass terms
$m^{ij}Q_iQ_j$ since the apparent masses may be absorbed in the shifts of the
singlets $Z^{ij}$.

It is straightforward to generalize the above model to
an Sp$(N)$ gauge theory
\refmark{\Pou}
with $2(N+1)$ chiral superfields
in the $2N$ representation.
Here we adopt a notation ${\rm Sp}(1) = {\rm SU}(2)$.

These vector-like models might serve as a supersymmetry-breaking mechanism
in the hidden
\refmark{\Iza}
or visible sector.


\refout

\bye